\documentclass[aip,jcp,a4paper,reprint,nofootinbib]{revtex4-1}

\usepackage[version=3]{mhchem}
\usepackage{braket}
\usepackage{graphicx}
\usepackage{siunitx}
\usepackage{amsmath}
\usepackage{amsfonts}
\usepackage{amssymb}
\usepackage{upgreek}
\usepackage{multirow}

\usepackage{booktabs}

\begin{document}
\title{Theoretical Prediction of Magnetic Exchange Coupling Constants 
from Broken-Symmetry Coupled Cluster Calculations}

\author{Henry F. \surname{Schurkus}}
\affiliation{
Division of Chemistry and Chemical Engineering, 
California Institute of Technology, Pasadena, CA 91125
}
\author{Dian-Teng \surname{Chen}}
\author{Hai-Ping \surname{Cheng}}
\affiliation{
Department of Physics, 
University of Florida, Gainesville, FL 32611
}
\author{Garnet Kin-Lic \surname{Chan}}
\affiliation{
Division of Chemistry and Chemical Engineering, 
California Institute of Technology, Pasadena, CA 91125
}
\author{John F. \surname{Stanton}}
\affiliation{
Department of Chemistry, 
University of Florida, Gainesville, FL 32611
}

\begin{abstract}
  Exchange coupling constants ($J$) are fundamental to the understanding of
  spin spectra of magnetic systems.
  Here we investigate the broken-symmetry (BS) approaches of Noodleman and Yamaguchi in conjunction with coupled cluster (CC)
  methods to obtain exchange couplings.
  $J$ values calculated from CC in this fashion converge smoothly towards the FCI result with
  increasing level of CC excitation.
  We compare this BS-CC scheme to the complementary
  EOM-CC approach on a selection of bridged molecular cases and
  give results from a few other methodologies for context. 
\end{abstract}
\maketitle

\section{Introduction}

The energy level structure of spin states is fundamental to the description of
magnetism in molecules and materials.
For molecules with localized spins on different atoms, the
low-energy spin-states
can often be qualitatively understood in terms of the
phenomenological Heisenberg model\cite{Dirac1926,Heisenberg1928,VanVleck1932}
\begin{equation}
  H = -2 \sum_{A,B>A} J_{AB} ~ \textbf{S}_A \cdot \textbf{S}_B
  \label{Heisenberg}
\end{equation}
\noindent where $A$ and $B$ index the ``spin centers''.
The Heisenberg model is completely parametrized by
the magnetic exchange coupling
constant, $J_{AB}$, for each spin interaction $A-B$.

Estimating the exchange coupling, and its geometric dependence, is
complicated by the fact that the underlying mechanism of spin-interactions
is a multi-electron process, such as Anderson super-exchange;\cite{Anderson1959}
furthermore, the low spin electron configurations that often appear
in such investigations are a formidable challenge to quantum chemical methods.
The most commonly-used approach involves calculations with density functional
theory (DFT).
Although DFT is ill-suited to describe 
eigenstates of the Heisenberg model, which possess
multireference character  arising from
the largely independent spin orientations of the different centers,
correctly parametrizing the model only requires us to match the energies
of low-energy states, which need not be chosen as eigenstates.
Consequently, 
it is
commonly found that approaches based on broken-symmetry (BS) spin states
such as the ones proposed by Noodleman\cite{Noodleman1981} and Yamaguchi\cite{Yamaguchi1984}
can give estimates of $J$ that are qualitatively comparable to experimentally-extracted
values, even in cases where the exchange coupling arises due to
super-exchange.~\cite{Noodleman1986,Nishino1997,Caballol1997,Ruiz1999,Ruiz2004,Rudra2006,Comba2009,Pantazis2010} 
Still it is worthwhile to explore more
sophisticated 
approaches within electronic structure, as this potentially permits the
intrinsic Heisenberg energy level structure to be predicted with quantitative
accuracy.

Coupled cluster (CC) theory is often used to generate benchmark quality
descriptions of molecular properties.~\cite{Shavitt2009} 
Recently, Mayhall and Head-Gordon used spin-flip equation-of-motion (EOM) CC
methods to obtain exchange couplings,\cite{Mayhall2014} based on 
 using CC and EOM-CC to approximate the two eigenstates of highest and next-highest spin
described by the Heisenberg model. However, as mentioned, it is not necessary to 
target spin-eigenstates when parametrizing the Heisenberg model.
Here we adopt the broken-symmetry methods of Noodleman and Yamaguchi
in conjunction with coupled cluster theory 
to estimate the exchange parameters. We assess this broken-symmetry CC technique
in a variety of magnetically coupled small molecules and bridged transition metal dimers.

\section{Theory}

\subsection{Extracting exchange couplings}\label{sec:Curve}

Where used, the Heisenberg model is intended to describe the low-energy
spin excitations of the system, but such a description is necessarily approximate.
Thus the value of the exchange coupling depends in part
on the way in which it is extracted from data. Experimentally, 
values reported in laboratory studies are generally obtained by
fitting the measured magnetic susceptibility 
to predictions based on the Heisenberg model.

\begin{figure}
  \centering
  \includegraphics[width=.5\textwidth]{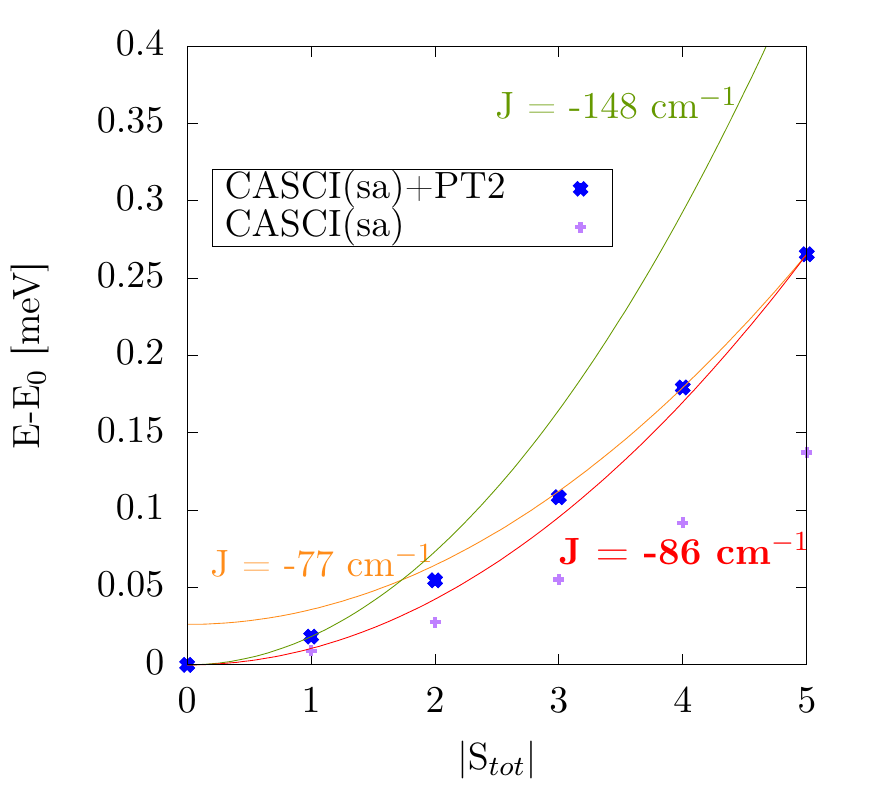}
  \caption{Theoretical spectrum of \ce{Fe2OCl6^{2-}} as obtained by
    CASSCF(10,10) and NEVPT2 with orbitals obtained via a
    spin-average over all spin-states.
    A simple Heisenberg model cannot exactly capture this
    spectrum and fits of the Heisenberg model
    yield exchange couplings that vary by up to a factor of 2 depending on the
    chosen weighting of the states in the fit.
    Note that for single-reference methods
    $|\textrm{S}_{tot}|$ can differ
    significantly from integer values.
    }
  \label{fig:Curve}
\end{figure}

Within theoretical approaches, we can easily illustrate the ambiguity 
in a system with only two spin centers like the ones studied
in this work, in which a single value of $J$ defines the Heisenberg model
completely.
For example, Fig.~\ref{fig:Curve} shows the spin ladder for \ce{Fe2OCl6^{2-}} 
computed from spin-averaged complete-active-space self-consistent field (CASSCF)
(10,10) orbitals~\cite{Sun2017} so as to treat all spin states on
equal footing and corrected by $n$-electron valence second-order perturbation
theory (NEVPT2)~\cite{Angeli2001,Angeli2001a,Angeli2002,Guo2016} to partially recover the lost correlation
from limiting the active space.
Choosing the two 
highest spin states (HS, HS-1) as is done in the procedure of Mayhall and Head-Gordon
gives a value of $J$ that is 71 cm$^{-1}$ smaller in magnitude than if the states
of lowest multiplicity are used, a discrepancy which is comparable to the J values
themselves. A least-squares fit to all states yields -85 cm$^{-1}$, which
is within 10\% of the former value, and even closer to that obtained when the
lowest spin and highest spin (LS and HS, respectively) states are selected. 
Note that strong $S$ dependence of $J$ in these fits does not necessarily mean that
the Heisenberg model is a poor approximation for the molecule itself, because
the quality of the theoretical approximations themselves depends on the spin state. 
Thus we see that, when giving a theoretical value for $J$ it is important to specify
which states were used to compute it, which we do in our work below.

Finally, we stress that it is not necessary, nor always desirable, to fit the exchange parameters
of the Heisenberg model to theoretical calculations of spin eigenstates. The basis
of an effective model is that there exists a space of low-energy states where the matrix elements
of the model Hamiltonian and the \textit{ab initio} Hamiltonian agree, but one is free to choose
any rotation within this space to characterize the model parameters. While fitting to
eigenstates is convenient, it is undesirable if the theoretical approach incurs a large
error for such states. This is the rationale behind broken symmetry approaches, which we now discuss.

\subsection{Broken symmetry approach to J couplings}

One of the earliest proposals to estimate exchange couplings from broken-symmetry wavefunctions
was given by Noodleman.~\cite{Noodleman1981}
His popular method
 computes magnetic 
exchange coupling constants using broken symmetry unrestricted Hartree-Fock (BS-UHF) solutions for low-spin states
\begin{equation}
  J = \frac{-[E(\textrm{HS}) - E(\textrm{BS})]}{s_{max}^2}, 
  \tag{2, "Noodleman"} \stepcounter{equation}
  \label{Noodleman}
\end{equation}
where $E(\textrm{BS})$ is the energy of the low-spin solution, $E(\textrm{HS})$ is 
the high-spin energy, and $s_{max}$ is the total spin of the high-spin
state. This assumes that the broken symmetry state is an equal mixture of the lowest
and highest spin states, which is strictly valid only for broken symmetry determinants
with two $s=1/2$ centers in the weak overlap limit.

A more general approach was suggested by Yamaguchi, originally for DFT calculations.\cite{Yamaguchi1984} That approach, and its correspondence to that
of Noodleman (which is also today applied with DFT calculations),\cite{Noodleman1981} can be developed as follows.  Consider two coupled 
spins ${\bf S}_A$ and ${\bf S}_B$, for which the resultant spin is
\begin{eqnarray}
{\bf S}_{tot}^2 = {\bf S}_A^2 + {\bf S}_B^2 + 2 {\bf S}_A \cdot {\bf S}_B.
\end{eqnarray}
Using the definition of $J_{AB}$, Eq.~\ref{Heisenberg}, the
energy of a given state $\psi$ (not necessarily an eigenstate) is 
\begin{eqnarray}
E(\psi) = -J_{AB} 
\left [ \langle {\bf S}_{tot}^2 \rangle - \langle {\bf S}_A^2 \rangle - \langle {\bf S}_B^2\rangle  \right ],
\end{eqnarray}
which can be used to determine $J_{AB}$ by using energies of any two
states
$\psi_1$ and $\psi_2$, {\it viz.}
\begin{eqnarray}
J_{AB} = \frac{E(\psi_1) - E(\psi_2)}{ \braket{\textbf{S}^2}_{\psi_2}  -
\braket{\textbf{S}^2}_{\psi_1}}.
\end{eqnarray}
Typically, one chooses $\psi_1$ to be an approximation to the HS state, which
is usually close to a spin eigenfunction with most methods.
For the case under consideration then,
one can obtain the specific form of the Yamaguchi formula
by inserting the HS (T) and BS (S) energies and spins
\begin{equation}
  J_{AB} = \frac{E(\psi_T) - E(\psi_S)}{ \braket{\textbf{S}^2}_{\psi_S} -
  \braket{\textbf{S}^2}_{\psi_T}} ~.
  \tag{8, "Yamaguchi"} \stepcounter{equation}
  \label{Yamaguchi}
\end{equation}

For two uncoupled spins, the broken-symmetry UHF singlet solution is roughly
``half-singlet'' and ``half-triplet'', so that $\braket{\textbf{S}^2}_{BS} \sim 1$, the equality of which
recovers the Noodleman formula with $s_{max}=1$
provided the high-spin wavefunction is a spin eigenfunction.  
Similarly, for the desired broken symmetry solution in which
all unpaired $\alpha$ spins are on one center, and all unpaired $\beta$ 
on the other, it
can be shown that $\langle {\bf S}^2 \rangle_{BS} = s_{max}$, so that the denominator
of the Yamaguchi formula reduces to
\begin{eqnarray}
s_{max} - s_{max} (s_{max} + 1 ) = -s_{max}^2,
\end{eqnarray}
which serves to show the correspondence between the Yamaguchi and Noodleman
equations.

The advantage of the Yamaguchi formula is that it can be applied to any
wavefunction for the low-spin state, approximate or exact,
while the Noodleman formula
(at least in the sense of the correspondence illustrated above) applies
only when the broken-symmetry wavefunction is used in its unadulterated
form, {\it i.e.} at the SCF (or Kohn-Sham DFT) level of theory.
The accuracy of the Yamaguchi formula then
depends on how completely the low-spin state is contained in the linear-span
of spin eigenstates that form the model space of the Heisenberg model, and how well
the theoretical method captures the expectation value of the energy in such a state.
It has been recognized that
coupled-cluster (CC) calculations based on broken-symmetry reference 
functions are an expedient way to obtain reasonably accurate energies 
in many situations qualitatively described by low-spin electronic 
configurations,\cite{Saito2011} such as in homolytic bond-breaking and some transition
states (similar strategies are followed in broken-symmetry DFT, which is
often referred to as broken-symmetry unrestricted Kohn-Sham theory (BUKS)).
As the expectation value of $\textbf{S}^2$ is easily 
calculated for coupled-cluster wavefunctions,\cite{Stanton1994a} it is thus worthwhile to explore 
the Yamaguchi formula to calculate magnetic exchange 
coupling constants using broken-symmetry CC wavefunctions, and such calculations form the
core of the work reported here.

\section{Illustrative Calculations}
$J$ values for a series of molecules with bridged spin centers will now be
presented,
comparing the BS-CC approach described above to the EOM-CC approach described
previously by
Mayhall and Head-Gordon. For reference, we will also give results obtained by
the most commonly used approach, evaluating the Noodleman formula with DFT
orbitals, and a few other methods.

\subsection{Computational Details}\label{sec:computational}
All calculations were carried out in the cc-pVDZ basis\cite{Dunning1989,Woon1993} unless specified
otherwise, or in plane-wave bases where denoted by PW.
PBE, HF, CAS, EOM, and CCSD(T) results were generated with
\texttt{pyscf}\cite{Sun2018,Sun2017,Guo2016,Github:pyscf}.
Coupled cluster results beyond CCSD(T) were
generated with \texttt{CFOUR}\cite{Stanton}
and the \texttt{MRCC} program of K\'allay.\cite{Kallay2019,Rolik2013a} PW-DFT
results were generated in \texttt{VASP}\cite{Kresse1996,Kresse1996a} for a
simple check of the robustness of the procedures to computational basis. 

For the Gaussian orbital calculations, 
orbitals were first obtained via a restricted open-shell
calculation (ROKS/ROHF) for the HS state. Guess orbitals for the LS solution
were derived by localizing the singly occupied space of the ROKS/ROHF solution
and assigning $\alpha$ and $\beta$ occupancies to them, which were subsequently converged
to the BS-UKS/BS-UHF ground state.
In addition, HS UKS orbitals were computed, taking care to break
spatial symmetry when present in order to obtain the lowest energy solution.

For the plane-wave calculations
projector-augmented-wave (PAW)
pseudopotentials\cite{Blochl1994,Kresse1999} 
were employed with 
a plane-wave cutoff energy of 500 eV and an energy threshold for self-consistency of  $10^{-6}$ eV.

Correlated wavefunction calculations were carried out starting from the Gaussian orbital mean-field solutions.
UCCSD calculations were based on the corresponding (HS/LS) HF solution keeping
all core orbitals frozen. For the
BS approach, the BS-UHF orbitals were used. For the EOM approach, ROHF
orbitals were used since this allowed for easier convergence of the EOM
amplitudes. To initialize the EOM eigenvectors into the correct space, a small
EOM calculation was carried out 
freezing all but the singly occupied orbitals. The singles amplitudes from
this calculation were then taken as an initial guess for the eigenvectors in the full space EOM calculation.
Preliminary testing showed $\textbf{S}^2$ values computed by CCSD
and CCSD(T) to be similar. To avoid large memory requirements for the larger
systems, $\textbf{S}^2$ values computed by CCSD were used for CCSD(T) as
well. 

CASCI calculations were performed using ROHF/ROKS orbitals, choosing all
singly occupied orbitals as the active space.
Further CASCI calculations were performed using 
orbitals determined from  spin-averaged CASSCF calculations over the same space, weighting the HS and LS
state equally (CASCI(sa)). Second-order perturbative corrections were calculated
for all cases separately via NEVPT2 (denoted ``+PT2'' below). Because both CASCI and NEVPT2 used a spin-adapted
implementation, the spins appearing in the Yamaguchi formula for these methods are equivalent to the spins of the eigenstates.

\subsection{Comparison to the full configuration interaction
limit}\label{sec:HHeH}

\begin{table}
  \begin{tabular}{lccc}
    & \multicolumn{1}{c}{Noodleman} 
    & \multicolumn{1}{c}{Yamaguchi}
    & $\textbf{S}^2_\textrm{LS}$\\[2mm]

    \underline{\ce{H-He-H}} \\[2mm]
    CCSDTQ          & -1126 & \textbf{-563} & 0.0000\\
    CCSDT           & -1120 & ~-560 & 0.0001 \\
    CCSD(T)         & ~-875 & ~-534 & --- \\
    CCSD            & ~-761 & ~-464 & 0.6873 \\\hline
    HF  (ROHF/BS-UHF) & ~-536 & ~-530 & \multirow{2}{*}{\Big\} 0.9883} \\
    HF  (UHF/BS-UHF) & ~-450 & ~-444 \\
    PBE (ROKS/BS-UKS) & -1092 & -1035 & \multirow{2}{*}{\Big\} 0.9450} \\
    PBE (UKS/BS-UKS) & -1036 & ~-981 \\[5mm]

    \underline{\ce{[H-F-H]^-}} \\[2mm]
    CCSDTQPH        & -2550& \textbf{-1275} & 0.0000\\
    CCSDTQP         & -2548 & -1274 & 0.0000\\
    CCSDTQ          & -2520 & -1260 & 0.0004\\
    CCSDT           & -2288 & -1164 & 0.0361\\
    CCSD(T)         & -1577 & -1202 & --- \\
    CCSD            & -1246 & ~-950 & 0.6880\\\hline
    HF  (ROHF/BS-UHF) & ~-803 & ~-789 & \multirow{2}{*}{\Big\} 0.9823}\\
    HF  (UHF/BS-UHF) & ~ +99 & ~ +97 \\
    PBE (ROKS/BS-UKS) & -3748 & -2702 & \multirow{2}{*}{\Big\} 0.6130}\\
    PBE (UKS/BS-UKS) & -3474 & -2500 \\
  \end{tabular}
  \begin{tabular}{lrr}
    \vspace{5mm}\\\hline 
    & \ce{H-He-H} & \ce{[H-F-H]-} \\\hline
    EOM-CCSD       &  -554 & -1178 \\
    PW-LDA         & -1435 & -2048 \\
    PW-PBE         &  -978 & -1589 \\
    PW-B3LYP       & -1240 & -1971 \\
    PW-SCAN        & -1009 & -1450 \\
    CASCI(PBE)     &  -399 & +5523 \\
    CASCI(PBE)+PT2 &  -537 & -5377 \\
    CASCI(HF)      &  -421 &  -126 \\
    CASCI(HF)+PT2  &  -519 &  -940 \\
    CASCI(sa)      &  -508 &  -425 \\
    CASCI(sa)+PT2  &  -536 &  -906 \\\hline
\end{tabular}
\caption{$J$ coupling constants in cm$^{-1}$ for \ce{H-He-H} and \ce{[H-F-H]-}
  across different methods. FCI quality results are highlighted in bold. For
  mean-field methods the HS and LS method are given respectively in parenthesis.
  Since $\textbf{S}^2$ are not computed in VASP, the Noodleman formula is used
  for all PW results. In all PW calculations the HS state is described by UKS.
}
\label{tab:HHeH}
\end{table}

We first look at two cases which
can be solved 
effectively exactly (i.e. full CI quality results are available)
in Table~\ref{tab:HHeH}.
Both model systems comprise two
spin-$\frac{1}{2}$ centers coupled via super-exchange into a singlet and a triplet.
Both structures are centrosymmetric molecules comprising two hydrogen atoms bridged by a central
closed shell atom (X=\ce{He}, R(H-He)=1.5\AA, and \ce{F^-}, R(H-F)=2\AA{}).

Applying the Yamaguchi formula, the series of CC methods converges smoothly to the FCI limit. 
CCSDTQ is exact for \ce{H-He-H} and CCSDTQPH can already be seen as almost
converged for \ce{[H-F-H]-}, where FCI requires octuple excitations.
In routine chemical practice, however, calculations beyond CCSD(T) are rarely feasible.
It is encouraging that $J$ values obtained with BS-CCSD and EOM-CCSD are comparable in both
cases and in good agreement with the exact limit. 
Specifically, they are
considerably closer than the traditionally
used Noodleman approaches with mean-field methods. 
As the coupled cluster series approaches the exact limit, the corresponding
$\braket{\textbf{S}^2}_\textrm{LS}$ values have to decay from the broken-symmetry
value of the reference determinant to the spin eigenfunction value of zero.
Therefore, applying
the Noodleman formula with coupled cluster energies with increasing excitation level must converge to the wrong
result since it does not take this effect into account.
Since the deviation of the spin of the LS state
from the broken-symmetry value is already substantial within the CCSD
description,
especially for H-He-H ($\textbf{S}^2 = 0.362$), it is critical for the BS-CCSD
approach to employ the
Yamaguchi and not the Noodleman formula to correct for the non-zero $\textbf{S}^2$ value. Without any
correction, one would obtain only $J=$-437 cm$^{-1}$ even with CCSD(T), while
the Noodleman formula would drastically overshoot (see Table~\ref{tab:HHeH}). This difference between
the Noodleman and Yamaguchi equations does not occur within the mean-field description
for which the Noodleman approach was originally intended, as the BS $\textbf{S}^2$ 
value (0.998) is quite close to the ideal value of 1.
We will study in
Section~\ref{sec:TMcomplexes} how important this difference is in real molecular systems.
Surprisingly, for the \ce{H-He-H} case, EOM-CCSD even outperforms BS-CCSD(T). We will
see in Section~\ref{sec:TMcomplexes} that this is not always the case in realistic
molecules.

While Noodleman
and Davidson originally suggested their equation for HF, it is often used
with density functionals instead. While the Noodleman ROHF/BS-UHF results for these
cases are only off by up to 37\%, the corresponding PBE results can be off by more than a
factor of two. Similar results are seen in both the Gaussian and PW
basis.

CASCI underestimates the magnitude of the coupling constant since the active space only
correlates the valence electrons of the two spin-centers and thus does not capture
the super-exchange mechanism. NEVPT2 treats the effect of all other
electrons perturbatively and recovers part of the missing
correlation. We find that NEVPT2 still underestimates the missing correlation
and therefore the magnitude of $J$, although it
outperforms all mean-field methods independent of whether the
Noodleman or Yamaguchi formula are used.

\subsection{Application to bridged transition metal dimers}\label{sec:TMcomplexes}

\begin{figure}
  \includegraphics[width=.5\textwidth]{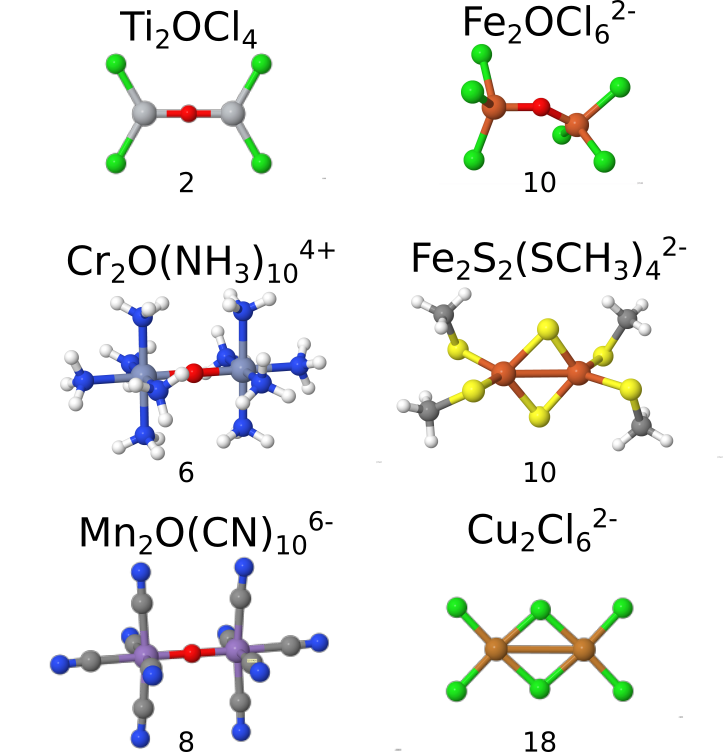}
  \caption{Bridged transition metal dimers for which J values are evaluated in
  Section~\ref{sec:TMcomplexes}. The total number of d-electrons is given below
each molecule.}
  \label{fig:Molecules}
\end{figure}

We next consider how these findings generalize to realistic bridged transition metal
dimers with varying numbers of d-electrons (Figure~\ref{fig:Molecules}).

\begin{table}
  \begin{tabular}{lr>{\tiny}lr>{\tiny}lr}
    & $E(\textrm{HS})$ & ($\textbf{S}^2_\textrm{HS}$) &
    $E(\textrm{low})$ &
    ($\textbf{S}^2_\textrm{
    low}$) & J\\\hline
    & \multicolumn{1}{r}{a.u.} & 
    & \multicolumn{1}{r}{a.u.} & 
    & \multicolumn{1}{r}{cm$^{-1}$}\\[4mm]

    \textbf{\ce{Ti2OCl4}}
    & \multicolumn{1}{l}{-3611\ldots} \\
    EOM-CCSD & .533061 &  & .533283 & & -24\\
    BS-CCSD & .533126& (2.002079) & .533267 & (0.999454)   & -31 \\
    BS-CCSD(T) & .583558 & & .583748 & & -42 
    \\\multicolumn{6}{c}{}\\

    \textbf{\ce{Cr2O(NH3)10^{4+}}}
    & \multicolumn{1}{l}{-2725\ldots} \\
    EOM-CCSD & .426271 & & .430946 & & -171 \\
    BS-CCSD & .426729& (12.008289) & .433326 & (2.910380) & -159 \\
    BS-CCSD(T) & .514911 & & .522934& & -194
    \\\multicolumn{6}{c}{}\\

    \textbf{\ce{Mn2O(CN)10^{6-}}}
    & \multicolumn{1}{l}{-3300\ldots} \\
    BS-CCSD & .159211& (20.040980) & .189390& (3.805756) & -408 \\
    BS-CCSD(T) & .363566& & .375919& & -167
    \\\multicolumn{6}{c}{}\\

    \textbf{\ce{Fe2OCl6^{2-}}}
    & \multicolumn{1}{l}{-5359\ldots} \\
    EOM-CCSD & .119664& & .125261 & & -123 \\
    BS-CCSD & .120087& (30.003902) & .131194& (4.796592) & -97 \\
    BS-CCSD(T) & .178681 & & .192750 & & -123
    \\\multicolumn{6}{c}{}\\

    \textbf{\ce{Fe2S2(SCH3)4^{2-}}}
    & \multicolumn{1}{l}{-5071\ldots} \\
    BS-CCSD & .020339 & (30.014220) & .038130& (4.699074) & -154 \\
    BS-CCSD(T) & .107047 & & .129098& & -191
    \\\multicolumn{6}{c}{}\\

    \textbf{\ce{Cu2Cl6^{2-}}}
    & \multicolumn{1}{l}{-6037\ldots} \\
    EOM-CCSD & .368900 & & .368900& & -0 \\
    BS-CCSD & .369189 & (2.001204) & .369169& (0.998728) & 4 \\
    BS-CCSD(T) & .414833 & & .414886 & & -12 \\
  \end{tabular}
  \caption{State-specific absolute energies and $J$ values computed
    via EOM-CCSD, BS-CCSD, and BS-CCSD(T). For EOM ``low'' denotes the HS-1 state,
    for BS the broken-symmetry LS state. The EOM HS energy is obtained from
    ROHF orbitals, the BS HS energies from UHF orbitals. The exact values for
    $\textbf{S}^2_\textrm{HS}$ and $\textbf{S}^2_\textrm{HS-1}$ were used for
    the evaluation of $J$ from EOM
    to reproduce the procedure used by Mayhall and Head-Gordon.\cite{Mayhall2014} For BS-CCSD(T)
  the $\textbf{S}^2$ values computed with BS-CCSD were used.}
  \label{tab:CCcomp}
\end{table}

The different CC approaches are contrasted in Table~\ref{tab:CCcomp}.
In line with \citet{Mayhall2014}, for EOM-CCSD the $\textbf{S}^2$ values were not calculated but their idealized values
(assuming spin eigenstates for HS and HS-1) were used. All three methods--EOM-CCSD, BS-CCSD, and
BS-CCSD(T)--yield comparable results. As in the cases of \ce{H-He-H} and
\ce{[H-F-H]^-}, for all systems other than \ce{Mn2O(CN)10^{6-}} going from BS-CCSD to BS-CCSD(T) increases the magnitude of $J$.
In all these cases then, the FCI limit is probably slightly larger in magnitude
than the BS-CCSD(T) result. Given this assumption, BS-CCSD(T) performs best
across all molecules. There is no clear trend as to whether BS-CCSD or EOM-CCSD
performs better.

\begin{figure}
  \includegraphics[width=.5\textwidth]{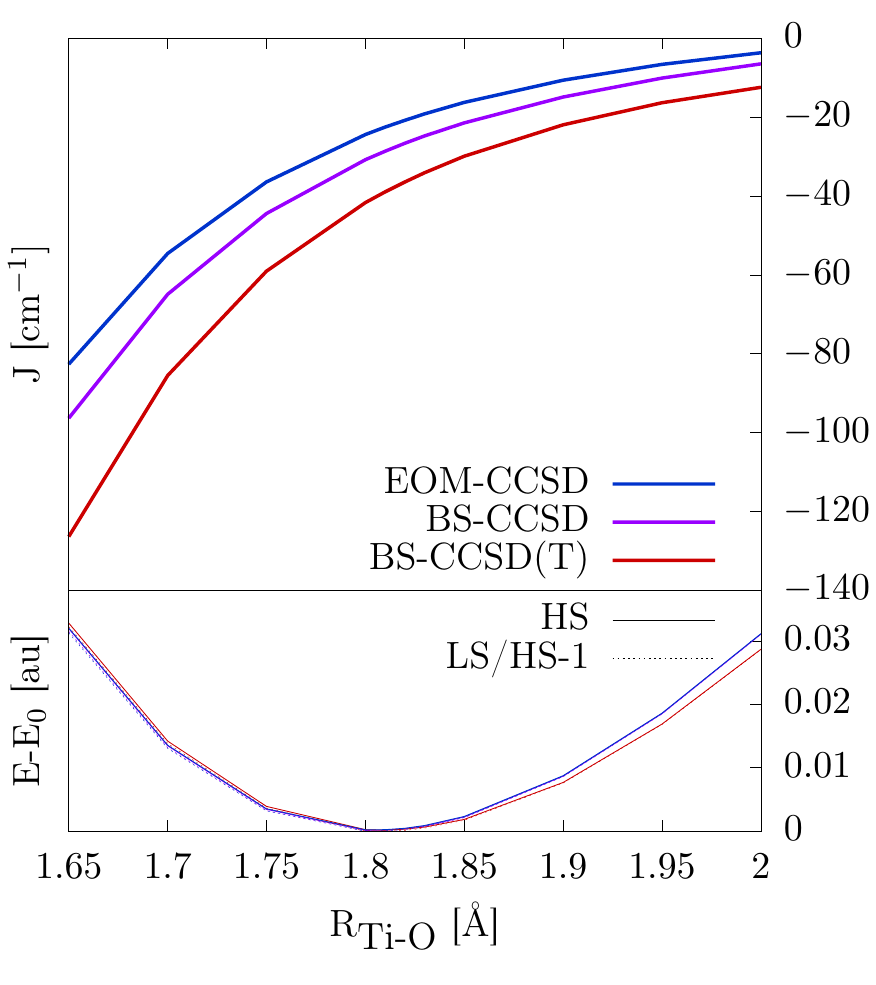}
  \caption{$J$ coupling constants (top) and HS and LS energies
    (bottom) with BS-CCSD, BS-CCSD(T) and EOM-CCSD (HS-1 instead of LS for the
    EOM calculation). Note that in the lower panel, the LS curves are almost directly under the HS curves
    due to the small size of the exchange coupling on this scale. For each of the three methods the energy curves have been
    shifted by their respective LS equilibrium energies, $\textrm{E}_0$. All
    three methods qualitatively agree near equilibrium, with BS-CCSD(T) starting to visually differ at
    stretched distances. Nonetheless, the resulting J coupling constant distance dependence
    is similar in all methods even when the 
    absolute energies start to differ.}
  \label{fig:Ti2OCl4curve}
\end{figure}

All three methods are consistent even away from equilibrium geometry.
Figure~\ref{fig:Ti2OCl4curve} shows the energy curves for \ce{Ti2OCl4} with
respect to
symmetric stretching of the \ce{Ti-O} bond distance maintaining all other angles and
distances and the corresponding $J$ values. All methods
agree regarding the equilibrium distance and show $J$ to (properly) decay towards zero as
the bond is dissociated at similar rates.

\begin{table*}
  \begin{tabular}{l>{\bfseries}c >{\bfseries}cccccc}\vspace{3mm}
    {\tiny Spin Descr.} & \normalfont\tiny Yamaguchi & \normalfont\tiny Mayhall
    & \tiny Noodleman & \tiny Noodleman & \tiny Yamaguchi & \tiny Yamaguchi\\[-3mm]
    J [cm$^{-1}$] & CCSD(T) & EOM-CCSD & ROHF & UHF & UKS (PBE)
    & CASCI(sa)+PT2 & Experiment \\[2mm]
  \multicolumn{1}{l|}{\ce{Ti2OCl4}}         & -41 & -24 & -982 & +3 & -77 & -12 &
  \multicolumn{1}{|c}{---\footnote{no experimental result available. Geometry taken from \textcite{Hart1992} at R(Ti-O)=1.8 \AA.}}\\
  \multicolumn{1}{l|}{\ce{Cr2O(NH3)10^4+}}  
  & -194\footnote{\label{dzsv}$\textbf{S}^2$ values computed from analogous calculation with def-sv basis on the outer ligands due to memory limitations (energies full cc-pvdz)} 
                                                  & -171 & -424 & -60 & -356 & -89 &
  \multicolumn{1}{|c}{-225\footnote{from \textcite{Pedersen1972}. Geometry taken from \textcite{Harris2014} who reported J=-124 cm$^{-1}$.}} \\
  \multicolumn{1}{l|}{\ce{Mn2O(CN)10^6-}}   
  & -167$^{\tt\footnotemark[2]}$ 
  & ---\footnote{UCCSD did not converge due to the large ROHF$\rightarrow$UHF
  instability of the HS state ($\Delta$E=3.3 eV, $\Delta\textbf{S}^2$=0.74).} 
                                                        & -2966 & -1238 & -2085	& -2000 &
  \multicolumn{1}{|c}{-360\footnote{estimated from \textcite{Ziolo1974a} by
    fitting the measured magnetic moment to a two-site Heisenberg model. The
  experimental result may be smaller due to unknown amounts of
paramagnetic impurities in the sample.}}\\
  \multicolumn{1}{l|}{\ce{Fe2OCl6^2-}}      & -123 & -123 & -164 & -23 & -246 & -70 & 
  \multicolumn{1}{|c}{-112\footnote{from \textcite{Haselhorst1993}.
  Geometry taken from \textcite{Harris2014} who reported J=-117 cm$^{-1}$.}} \\
  \multicolumn{1}{l|}{\ce{Fe2S2(SCH3)4^2-}} 
  & -191$^{\tt\footnotemark[2]}$ 
  & ---\footnote{UCCSD did not converge due to large ROHF$\rightarrow$UHF
  instability of the HS state ($\Delta$E=1.9 eV, $\Delta\textbf{S}^2$=0.06).} 
                                                          & -433 & -54 & -850 & -111 &
  \multicolumn{1}{|c}{-148\footnote{from \textcite{Gillum1976} for the synthetic analog
      \ce{Fe2S2(S2\it{-o-}xyl)2^2-}. Geometry taken from \textcite{Sharma2014} who
report J=-236 cm$^{-1}$.}}\\
  \multicolumn{1}{l|}{\ce{Cu2Cl6^2-}}       & -12 & 0 & -1720 & +24 & -140 & -4 &
  \multicolumn{1}{|c}{-19\footnote{from \textcite{Maass1967}. Geometry taken
  from \textcite{Willett1963a}.}}\\
\end{tabular}
\caption{Comparison of J values computed by broken-symmetry
  CCSD(T) to those obtained with EOM-CCSD for a series of molecules depicted in
  Figure~\ref{fig:Molecules}. HF and PBE results from UHF and UKS calculations as
  well as CASCI(sa)+PT2, and experimental results are given for reference.
  Results obtained from the commonly used Noodleman approximation with ROHF
  and BS-UHF energies as suggested by Hart
  \textit{et al.}\cite{Hart1992} are also given.
  In the procedure described in Section~\ref{sec:computational} for the evaluation
  of EOM-CCSD, a large ROHF$\rightarrow$UHF instability leads to convergence
  problems for the CCSD calculation underlying EOM-CCSD in two cases (\ce{Mn2O(CN)10^6-}: $\Delta$E=3.3 eV,
  $\Delta\textbf{S}^2$=0.74; \ce{Fe2S2(SCH3)4^2-}: $\Delta$E=1.9 eV,
  $\Delta\textbf{S}^2$=0.06). In all other cases $\Delta$E is 0.4 eV or less and
  $\Delta\textbf{S}^2$ at most 0.03. ($\Delta$E and $\Delta\textbf{S}^2$ represent
  the change in energy and squared spin between ROHF and UHF).
  }
\label{tab:TMdimers}
\end{table*}

We contrast BS-CCSD(T)
and EOM-CCSD 
with results from mean field calculations as well as CASCI(sa)+PT2 
and experiment in Table~\ref{tab:TMdimers}. Both CC approaches are broadly consistent
with experimental results in all cases. 
From this, one can surmise that CC methods provide reliable results which can be
used to compare with other methods.

While a rigorous benchmark of different mean-field approaches is beyond the scope
of this study, the following
deserves mention: Hart \textit{et al.}\cite{Hart1992} had concluded from studying H-He-H, \ce{[H-F-H]^-}, and \ce{Ti2OCl4}
that the Noodleman formula with HF was performing better when using restricted
open-shell rather than unrestricted HS energies. 
While we can reproduce this effect for the same molecules (in fact, for two of
them using unrestricted HS orbitals even yields the wrong sign),
this seems not to be true in general.
In all cases involving transition-metal systems, the mixed
ROHF-UHF approach tends to vastly overestimate the magnitude of the coupling
constant.

HF results are off drastically in many cases, regardless of the orbitals
and formula chosen (Noodleman or Yamaguchi), even as much as an order of magnitude in
the case of \ce{Mn2O(CN)10^{6-}}. The same can be said for PBE results.
That the BS-CCSD(T) result is obtained from the same BS-UHF orbitals
as the UHF $J$ values indicates that the rather poor results for
the other methods reflect true shortcomings of those
methods in the context of these applications. This is even true for
CASCI(sa)+PT2 which, apart from this case, follows
the same behavior as discussed previously.

\begin{table}
  \begin{tabular}{l|cc|cc||c}
    & \multicolumn{2}{c|}{Noodleman} & \multicolumn{2}{c||}{Yamaguchi} & EOM \\
     & CCSD(T) & CCSD & CCSD(T) & CCSD & CCSD\\
    \hline
    \ce{Ti2OCl4} & -42 & -31 & -42 & -31 & -24 \\
    \ce{Cr2O(NH3)10^4+} & -196 & -161 & -194 & -159 & -171 \\
    \ce{Mn2O(CN)10^6-} & -169 & -414 & -167 & -408 &  \\
    \ce{Fe2OCl6^2-} & -124 & -98 & -123 & -97 & -123 \\
    \ce{Fe2S2(SCH3)4^2-} & -194 & -156 & -191 & -154 &  \\
    \ce{Cu2Cl6^2-} & -12 & 4 & -12 & 4 & 0 \\
  \end{tabular}
  \caption{Comparison of $J$ coupling constants computed from
    CCSD and CCSD(T) energies and consistently evaluated $\textbf{S}^2$ values
    (Yamaguchi) and with theoretical HS/BS $\textbf{S}^2$ values in comparison
    (Noodleman). EOM-CCSD values are given for reference.
  }
  \label{tab:HFspin}
\end{table}

One interesting finding in this study concerns the values of
$\braket{\textbf{S}^2}$ for the BS-CC
wavefunctions. While the small systems that are treated in
Section~\ref{sec:HHeH} are
such that correlation at the CCSD level acts to significantly reduce the LS
$\braket{\textbf{S}^2}$ value from the near-unity value of the reference determinant, it turns out
that this is not true for the transition metals, where the correlation
contribution to $\braket{\textbf{S}^2}$ is rather small. It seems that in these cases, the
electrons within the spin-centers are being more ``correlated'' than are the
interactions between electrons on different spin centers. Because of this, it
is apparent that the simple Noodleman equation -- which does not require the
(somewhat expensive) calculation of $\braket{\textbf{S}^2}$ can be applied in conjunction with the
BS-CC wavefunctions.
We verify this in
Table~\ref{tab:HFspin} and find that indeed this approach yields results almost
identical to full BS-CCSD and BS-CCSD(T) respectively. It is important to note
that this simpler approach appears to work well in practice (with less than
triples excitations). As discussed previously, however, it is apparent that as one converges the level of CC
excitations in these molecules, $\braket{\textbf{S}^2}$ will tend to zero and this approach
has to eventually converge to the wrong limit. We have seen this in
Section~\ref{sec:HHeH}, where due to the small size of the molecules 
already CCSD resulted in significantly reduced $\braket{\textbf{S}^2}$ values.

\section{Conclusion}

This work demonstrates that a simple application of the broken-symmetry approach
for calculating magnetic exchange coupling constants in conjunction with coupled-cluster theory provides useful results
in practice. As such, this method complements recent
work by Mayhall and Head-Gordon that has used the spin-flip variant of
equation-of-motion coupled cluster theory.
The two approaches both rely on
fitting $J$ to two energies; the present method uses the highest and broken-symmetry lowest-spin
state, while the EOM-CC method uses the two highest spin states.
Note that there is no formal disadvantage to using broken symmetry states so long as the $\braket{\textbf{S}^2}$
values are computed for the states of interest, as in the formula of Yamaguchi. However, 
we have also shown that the simpler approach of Noodleman, which posits the value of $\braket{\textbf{S}^2}$ for
the broken-symmetry lowest-spin state, works as well in practice for many realistic molecules.

Computations by the present method are quite straightforward; one needs only to
find BS solutions to the self-consistent field equations to obtain a reference
single determinant, and to evaluate coupled-cluster energies
and (optionally) one- and two-electron density matrix elements ($\braket{\textbf{S}^2}$ is straightforwardly computed from
these) if the Yamaguchi formula is used.
In particular, one does not need to wrestle with converging
the EOM-CC equations or assigning spin states, which is not always 
straightforward.\cite{Mayhall2014} In our experience,
iterative solvers to the EOM equations can get stuck on higher energy solutions
unless initial guesses are constructed very carefully.
While we studied binary systems with only a single $J$ coupling in this work,
in many cases
one is interested in finding $J$ for each of multiple interactions in a
molecule separately.
BS coupled cluster methods can then
potentially be applied the same way as BS DFT -- by spin flipping into separate
configurations.

Calibrating other methods may be one of the main uses of more
accurate methods to determine exchange couplings. Since both the EOM and BS coupled cluster approaches agree broadly with
experiment and yield consistent results across all studied systems
even away from equilibrium, they represent a reliable gauge by which to assess the
accuracy of other methods.
This is especially valuable since we observe very different behavior for different classes of molecules.  
For example, we can confirm that in small model systems using the commonly
applied Noodleman formula with ROHF energies instead of UHF energies for the HS
state yields superior results as posited by Hart \textit{et al.}\cite{Hart1992} However, we observe the same not to be true for
the larger transition metal complexes.

In short, broken-symmetry coupled cluster theory provides a straightforward
methodology to predict magnetic exchange coupling constants, complementing
approaches that target spin-eigenstates, such as equation-of-motion coupled cluster
methods and complete-active-space techniques. 
It is especially reliable when employing the Yamaguchi equation, in which case it can cope with almost
arbitrary amounts of spin contamination. 

\section{Acknowledgements}

This work was supported as part of the Center for Molecular Magnetic Quantum Materials, an Energy Frontier Research Center funded by the
U.S. Department of Energy, Office of Science, Basic Energy Sciences under Award No. DE-SC0019330. Computations were performed at NERSC, UFRC,
and on the Caltech HPC cluster.
HFS acknowledges funding from the European Union's Framework Programme for Research and Innovation Horizon 2020 (2014-2020) under
the Marie Sk\l{}odowska-Curie Grant Agreement No. 754388 and from LMUexcellent as part of LMU Munich's funding as a University of Excellence within the framework of the German Excellence Strategy.  

\bibliography{library}

\end{document}